# VALIDATION TEST CASES FOR MULTI-PHYSIC PROBLEMS: APPLICATION TO MAGNETO-HYDRODYNAMIC NUMERICAL SIMULATIONS

**David CEBRON**
Ecole Centrale de Nantes
Rue de la Noë
44321 NANTES, France
david.cebron@eleves.ec-nantes.fr

**Jean-François SIGRIST**
Service Technique et Scientifique
DCNS Propulsion
44620 LA MONTAGNE, France
jean-francois.sigrist@dcnsgroup.com

**Vincent SOYER**
Service Technique et Scientifique
DCNS Propulsion
44620 LA MONTAGNE, France

**Pierre FERRANT**
Ecole Centrale de Nantes
Rue de la Noë
44321 NANTES, France

## Abstract

The present paper is concerned with the numerical simulation of magneto-hydrodynamic (MHD) problems with industrial tools. MHD has received attention some thirty to twenty years ago as a possible alternative in propulsion applications; MHD propelled ships have even been designed to that purpose. However such propulsion systems have been proved of low efficiency and fundamental researches in the area have progressively received much less attention over the past decades. Numerical simulation of MHD problem could however provide interesting solutions in the field of turbulent flow control. The development of recent efficient numerical techniques for multi-physic applications provide promising tool for the engineer for that purpose. In the present paper, some elementary test cases in laminar flow with magnetic forcing terms are analyzed; equations of the coupled problem are exposed and analytical solutions are derived in each case, highlighting the relevant non-dimensional number which drives the physics of the problem. Several analytical calculations are then proposed and discussed. The present work will serve as basis for validation of numerical tools (based on the finite element method) for academic as well as industrial application purposes.

## INTRODUCTION

Magneto-Hydro-Dynamic effects have received attention some thirty to twenty years ago as a possible alternative in propulsion applications; MHD propelled ships have even been designed to that purpose: the most famous example is the Japanese ship Yamamoto 1, which has been designed and build as prototype of MHD-propelled ship (see Fig. 1 featuring a photo of the MHD thrusters from Yamamoto 1). However such propulsion systems have been proved of low efficiency and fundamental researches in the area have progressively received much less attention over the past decades.

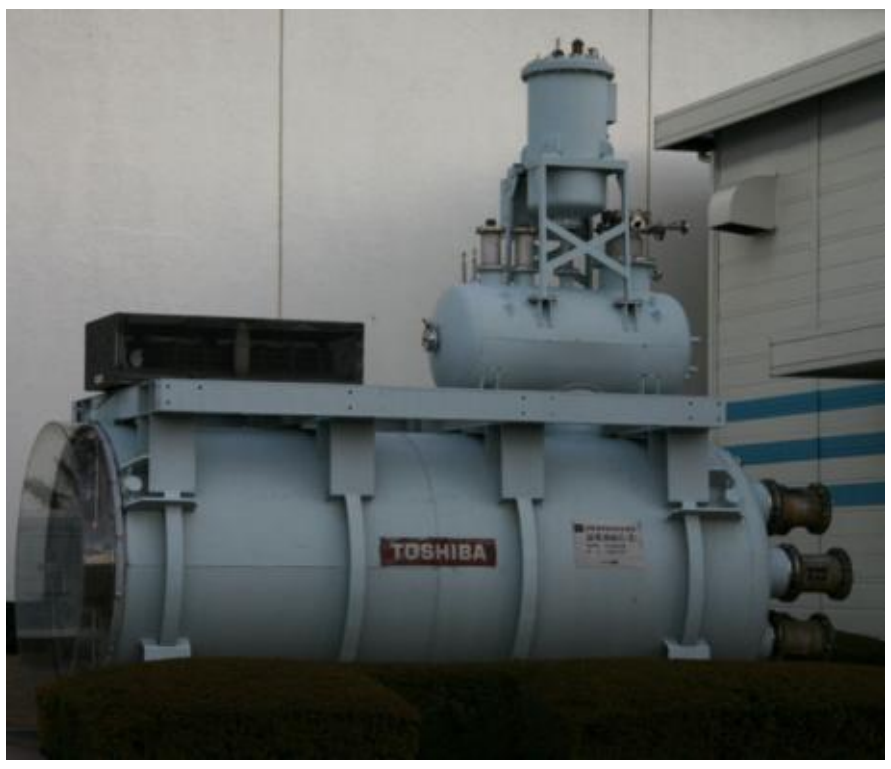

**Fig. 1. A MHD thruster from the experimental Japanese ship Yamato 1 at the Ship Science Museum in Odaiba, Tokyo.**

Numerical simulation of MHD problem could however provide interesting solutions in the field of turbulent flow control. The development of recent efficient numerical techniques for multi-physic applications provide promising tool for the engineer for that purpose. In the present paper, analytical test-cases in laminar flow with magnetic forcing terms are analyzed, namely the Hartman problem (section 1), the Couette problem (section 2) and the Rayleigh problem (section 3). An analytical solution is derived in each case[1] and the physic of the problem is discussed through the influence of a relevant non-dimensional number highlighted by the analytical expressions. As an illustration on MHD-based propulsion system, application of the Hartmann problem solution to an elementary propulsion nozzle is exposed. The present work will serve as basis for validation of numerical tools for multi-physic applications.

---

[1] *To the authors' knowledge, some of the presented analytical solutions have never been reported previously in the literature.*

## 1. HARTMANN PROBLEM

### 1.1. Problem definition

The classical Hartmann flow, first investigated by Hartmann in the 1930s, is probably the first validation test case to consider, see Fig. 2. The problem to be considered is the steady flow of an incompressible neutral but electrically conducting fluid in the positive $x$ direction, with a magnetic field $\mathbf{B}$, assumed to be uniform and constant, in the positive $z$ direction. Assuming an electrical conductivity infinite for the electrodes, we will neglect end effects ($L >> a$), secondary flows ($b >> a$) and the Hall effect, which allows to use a usual Ohm's law:

$$\mathbf{j} = \sigma(\mathbf{E} + \mathbf{v} \times \mathbf{B}) \tag{1}$$

where $\sigma$ is the electrical conductivity of the fluid.

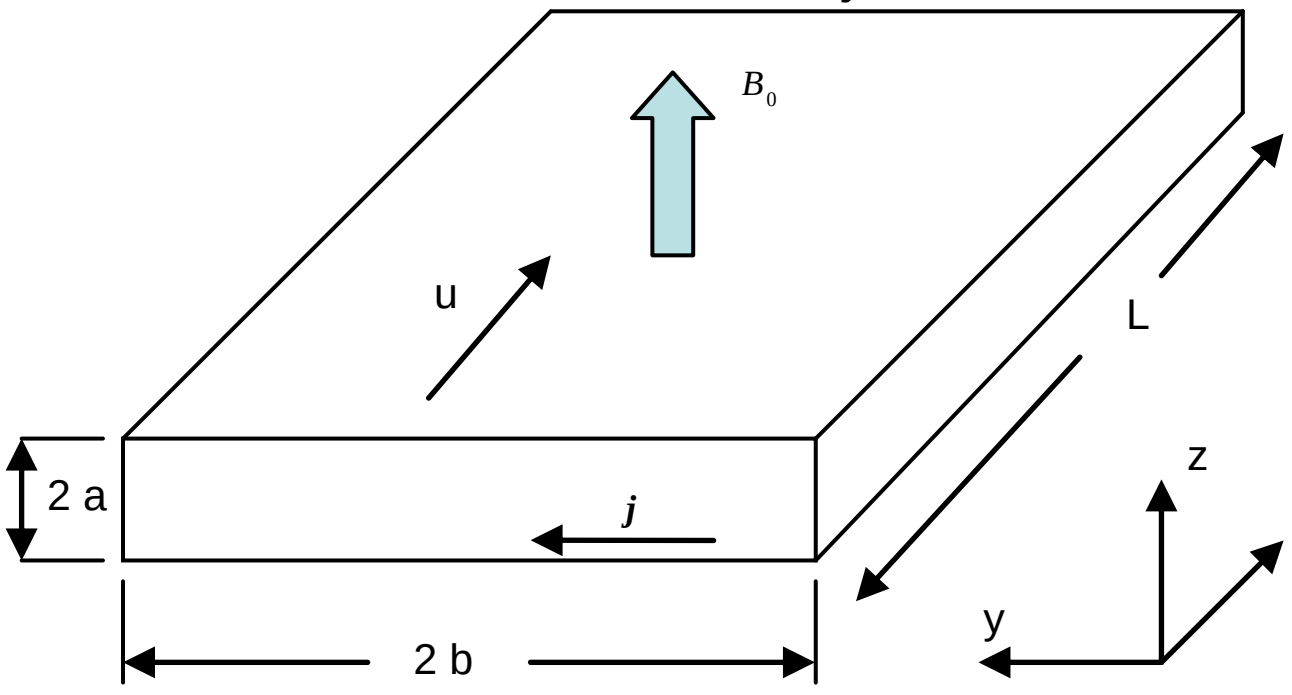

**Fig. 2. Channel geometry for Hartmann flow**

Except for the pressure $p$ and the temperature $T$, previous assumptions lead to variables functions of $z$ alone: $\mathbf{v} = (u_x, 0, 0)$, $\mathbf{j} = (0, j_y, 0)$, $\mathbf{B} = (b_x, 0, B_o)$, $\mathbf{E} = (0, E_o, 0)$ where $E_o$ and $B_o$ are constants.

We can notice that the problem geometry leads to $\nabla \cdot \mathbf{v} = 0$, so that, with the mass conservation equation,

$$\frac{\partial \rho}{\partial t} + \nabla \cdot (\rho \mathbf{v}) = \frac{D\rho}{Dt} + \rho \nabla \cdot \mathbf{v} = 0 \tag{2}$$

$\rho$ is constant (steady flow): the heat-transfer and fluid motion equations are then uncoupled. The general equations for this problem are then the classical equations of the MHD: Ohm's law (Eq. [1]), mass conservation Eq. [2]), Maxwell's equations Eqs. (3) to (6) bellow), equation of motion (Eq. [7] in which advective terms have been discarded) and energy equation (Eq. [8]) with viscous and ohmic dissipation, repeated here for convenience, assuming that viscosity $\eta$, electrical and thermal conductivity of the fluid, $\sigma$ and $\lambda$, are not temperature-dependant:

$$\nabla \times \mathbf{E} = -\frac{\partial \mathbf{B}}{\partial t} \tag{3}$$

$$\nabla \times \mathbf{B} = \mu \mathbf{j} + \varepsilon\mu \frac{\partial \mathbf{E}}{\partial t} \tag{4}$$

$$\nabla \cdot \mathbf{B} = 0 \tag{5}$$

$$\nabla \cdot \mathbf{E} = \rho_e \tag{6}$$

$$\rho \frac{\partial \mathbf{v}}{\partial t} = \mathbf{j} \times \mathbf{B} - \nabla p + \eta \Delta \mathbf{v} \tag{7}$$

$$\frac{\rho}{\lambda} \frac{De}{Dt} = \Delta T + \frac{\eta}{\lambda} \left\| \nabla \mathbf{v} + \nabla \mathbf{v}^T \right\|^2 + \frac{\mathbf{j} \cdot \mathbf{E}}{\lambda} + \frac{r}{\lambda} \tag{8}$$

where $\rho_e$ is the electric charge density, $e$ is the energy per unit mass of the fluid and $r$ a volumic source of heat.

### 1.2. Classical analytical solutions

Since the heat-transfer and fluid motion equations are uncoupled, it is possible to solve equations separately. Moreover the equations are linear and, following e.g. [1], the solution for the velocity profile $u_x$ and the mean velocity $\bar{u}$ is found to be (with boundary condition $u_x(\pm a) = 0$):

$$U = \frac{u_x}{\bar{u}} = \mathrm{H} \frac{\mathrm{ch(H)} - \mathrm{ch}(Z\mathrm{H})}{\mathrm{Hch(H)} - \mathrm{sh(H)}} \tag{9}$$

$$\bar{u} = \frac{\mathrm{H} - th(\mathrm{H})}{\mathrm{H}^3} \times \left( \mathrm{H}^2 \frac{E_o}{B_o} - \frac{a^2}{\eta} \frac{\partial p}{\partial x} \right) \tag{10}$$

where $Z = \frac{z}{a}$ and $\mathrm{H} = aB_o \sqrt{\frac{\sigma}{\eta}}$ is the so-called Hartmann number and $\mathrm{ch}(\bullet)$ and $\mathrm{sh}(\bullet)$ stand for hyperbolic cosine and sine, respectively (see figure 3 for plotted solutions).

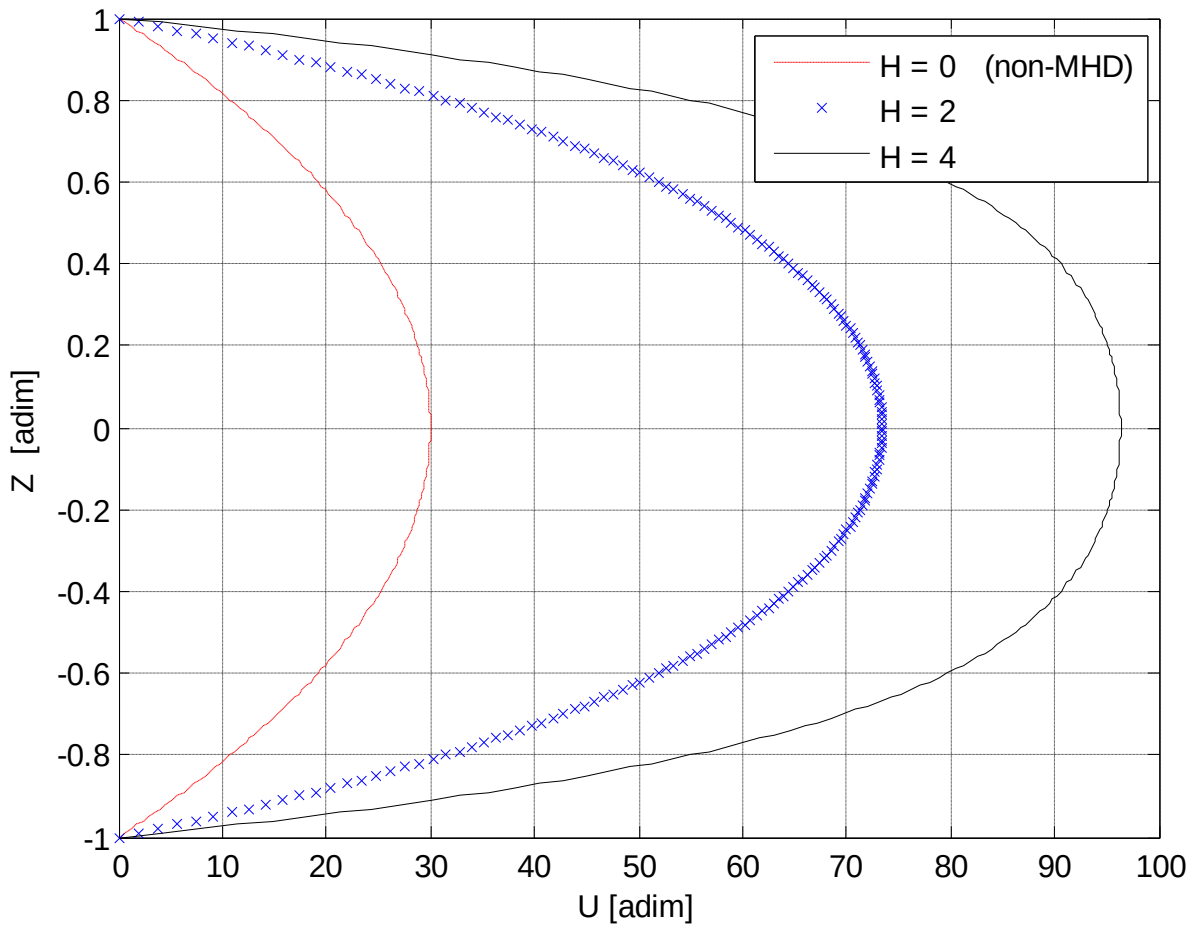

**Fig. 3. Hartmann velocity profiles ($K \cdot \bar{u} = 100$)**

It is interesting to know that a number of experimental investigations have provided excellent agreement with the previous solution (see e.g. [2]). Letting $K = \frac{E_o}{\bar{u} B_o}$ be the load factor of the flow and $J$ the net current flowing through the circuit per unit area, it is possible to deduce the reduced $\tilde{j}_y$ and the net density $J$ of currents:

$$\tilde{j}_y = \frac{j_y}{\sigma \bar{u} B_o} = K - \frac{u_x}{\bar{u}}, \quad J = (K-1)\sigma \bar{u} B_o \tag{11}$$

The electric field $E_o$ generated by a tension $\Omega$, is obtained here with Eq. (6) in the case of a parallel-wall channel $E_o = \frac{\Omega}{2b}$. The reduced induced magnetic field $\tilde{b}_x$ is derived from Eq. (4):

$$\tilde{b}_x = \frac{b_x}{B_o \mathrm{R_m}} = (K-1)Z + \frac{\mathrm{sh}(\mathrm{H}Z) - Z\,\mathrm{sh}(Z)}{\mathrm{H}\,\mathrm{ch}(\mathrm{H}) - \mathrm{sh}(\mathrm{H})} \tag{12}$$

where, as a consequence of symmetry of the problem, the boundary condition for $b_x$ is assumed to be $b_x = 0$. The magnetic Reynolds number $\mathrm{R_m}$ is defined here as $\mathrm{R_m} = \frac{a\bar{u}}{D_{\mathrm{mag}}}$, where $D_{\mathrm{mag}} = \frac{1}{\mu\sigma}$ is the magnetic diffusivity (see Eq. [49]).

Taking the state equation of an incompressible fluid $e = C_v T$ (with $C_v$ being the specific heat at constant volume), which is also the state equation of an ideal gas (second Joule's law), Eq. (8) can be written as: $\lambda \frac{\partial^2 T}{\partial z^2} + \eta\left(\frac{\partial u_x}{\partial z}\right)^2 + \frac{j_y^2}{\sigma} + r = 0$, hence the equation for temperature:

$$\frac{\partial^2 \theta}{\partial Z^2} = -\mathrm{P_r}\left[\left(\frac{\mathrm{dU}}{\mathrm{dZ}}\right)^2 + \mathrm{H}^2 \tilde{j}_y^2\right] - \tilde{r} \tag{13}$$

with the reduced temperature $\theta = \frac{C_v}{\bar{u}^2}(T - T_w)$ (which can be seen as the ratio between the thermal energy of the fluid and its kinetic energy), Prandtl number of the fluid $\mathrm{P_r} = \frac{\eta C_v}{\lambda}$, and the reduced source of heat $\tilde{r} = r\frac{C_v a^2}{\lambda \bar{u}^2}$. The boundary condition for $T$ is assuming to be: $T(\pm a) = T_w$, so that the analytical solution of Eq. (13) is :

$$\theta = C_1 \mathrm{P_r}\left[\frac{C_2^2}{2}(1 - Z^2) + \frac{1}{4}\left(\mathrm{ch}(2\mathrm{H}) - \mathrm{ch}(2\mathrm{H}Z)\right) + \frac{2C_2}{H}\left(\mathrm{ch}(\mathrm{H}) - \mathrm{ch}(\mathrm{H}Z)\right)\right] + \tilde{r}(1 - Z^2) \tag{14}$$

where constants $C_1$ and $C_2$ are given by $C_1 = \left(\frac{\mathrm{H}}{\mathrm{H}\,\mathrm{ch}(\mathrm{H}) - \mathrm{sh}(\mathrm{H})}\right)^2$ and $C_2 = \mathrm{H}(K\text{-}1)\mathrm{ch}(\mathrm{H})\ \text{-}K\mathrm{sh}(\mathrm{H})$.

The temperature profile according to Eq. (14) is represented by Fig. 4 for various values of the Harman number $\mathrm{H}$.

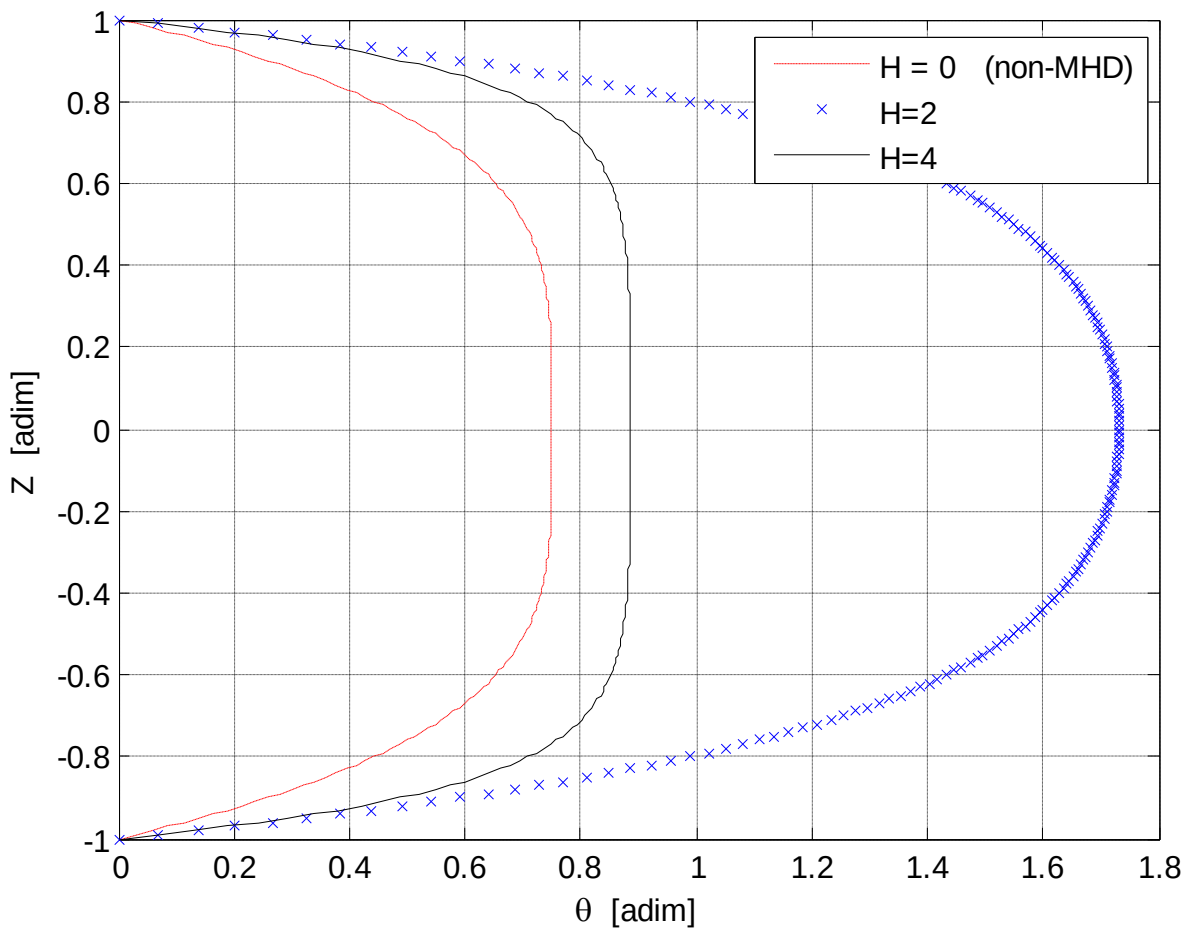

**Fig. 4. Hartmann flow temperature distribution ($\mathrm{K} \cdot \bar{\mathrm{u}} = 100$)**

It is interesting to know [11] that if the heat flux $q_w$ at the wall is independent of $x$ and the problem is assumed to be one-dimensional, then $T$ must be a linear function of $x$ :

$$T(x) = kx + g(z) \tag{15}$$

The boundary condition is $\frac{dg}{dz}(\pm a) = q_w$. Substituting Eq. (15) in Eq. (13) yields $g(z)$ and $\theta$, hence

$$k = \frac{q_w + \int_{-a}^{+a}\left(\eta\left(\frac{\partial u_x}{\partial z}\right)^2 + \frac{j_y^2}{\sigma}\right)dz}{\rho \bar{u} c_v a}.$$

It can be inferred that if $q_w = -\int_{-a}^{+a}\left(\eta\left(\frac{\partial u_x}{\partial z}\right)^2 + \frac{j_y^2}{\sigma}\right)dz$ then $k = 0$ and the previous problem solution is recovered since in this case, all the heat generated by viscous and Joule dissipation is transferred out of the channel. When $q_w$ is not constant or with a non-linear variation of $T$ along the channel wall, the problem become two-dimensional and rise many more difficulties [9].

### 1.3. Extension to an analytical solution with a thermal conductivity function linear of T

When the thermal conductivity is a function of $T$, the previous equation of energy (8) has to be modified in (16):

$$\frac{\rho}{\lambda}\frac{De}{Dt} = \Delta T + \frac{d\lambda_{(T)}}{dT}(\nabla T)^2 + \frac{\eta}{\lambda}\left\|\nabla \mathbf{v} + \nabla \mathbf{v}^T\right\|^2 + \frac{\mathbf{j}\cdot\mathbf{E}}{\lambda} + \frac{r}{\lambda} \tag{16}$$

The thermal conductivity of the fluid is assuming here to be a linear function of $T$ :

$$\lambda(T) = A_0 + A_1 T = A_0 + A_1(T_w + \theta\frac{\bar{u}^2}{c_v}) = \tilde{A}_0 + \tilde{A}_1\theta \tag{17}$$

Then, with $e = C_v T$ and the previous boundary condition for $T$, (16) can be solved with the Hartmann velocity profiles founded upper:

$$\theta_1 = -\frac{1}{\varepsilon_1} + \frac{1}{\varepsilon_1}\sqrt{1 + 2\varepsilon_1\left[C_1 \mathrm{P}_{r0}\left(\frac{C_2^2(1-Z^2)}{2} + \frac{\mathrm{ch}(2\mathrm{H}) - \mathrm{ch}(2\mathrm{H}Z)}{4} + \frac{2C_2\,\mathrm{ch}(\mathrm{H}) - \mathrm{ch}(\mathrm{H}Z)}{\mathrm{H}}\right) + \frac{\tilde{r}_0(1-Z^2)}{2}\right]} \quad (18)$$

where $\mathrm{P}_{r0} = \frac{\eta c_v}{\tilde{A}_0}$ is the Prandtl number at order 0 in $\varepsilon_1 = \frac{\tilde{A}_1}{\tilde{A}_0}$, and $\tilde{r}_0 = r\frac{c_v a^2}{\tilde{A}_0 \bar{u}^2}$, which allows to recover the previous solution, with a constant $\lambda$, at order 0 in $\varepsilon_1$, and then to obtain the gap between the two solutions at order 1 in $\varepsilon_1$: $E_{\max} = E(Z=0) \approx -\frac{\theta_0^2(Z=0)}{2}\varepsilon_1$.

### 1.4. Study of the Hartmann flow analytical solutions

First of all, it is interesting to remind the asymptotic expressions of previous solutions:

- $\mathrm{H} \ll 1$

$$U = \frac{u_x}{\bar{u}} = \frac{3}{2}(1 - Z^2) = u_{\max}(1 - \frac{z^2}{a^2}) \quad (19)$$

$$\bar{u} = \frac{1}{3}\left(H^2\frac{E_o}{B_o} - \frac{a^2}{\eta}\frac{\partial p}{\partial x}\right) \quad (20)$$

$$\theta = \frac{3}{4}\mathrm{P_r}(1 - Z^4) + \frac{\tilde{r}}{2}(1 - Z^2) \quad (21)$$

- $\mathrm{H} \gg 1$

$$U = \frac{u_x}{\bar{u}} \approx u\left(1 - e^{\mathrm{H}(|Z|-1)}\right) \quad (22)$$

$$\bar{u} \approx \frac{E_o}{B_o} - \frac{1}{\sigma B_o^2}\frac{\partial p}{\partial x} aE_o\sqrt{\frac{\eta}{\sigma}}\frac{1}{\mathrm{H}} \quad (23)$$

When $\mathrm{H}$ is assumed to be nil, the classical solution for no applied magnetic field is recovered, and when $\mathrm{H}$ becomes very large, the velocity profile is almost uniform. Moreover, at large $\mathrm{H}$, $\bar{u}\frac{E_o}{B_o}\xrightarrow[B_o \to \infty]{} 0$. Then, since $\bar{u}(B_o = 0) = 0$ without pressure gradient, there is a maximum for $\bar{u}$. In fact, this maximum comes from an equilibrium between the energy taken to accelerate the fluid and the energy lost in boundary layers, which can be associated with the efficiency ratio $r = \frac{D_v \Delta p}{UI} = \frac{\Delta p}{E_o} \times \frac{\bar{u}}{E_o - \bar{u}B_o}$. It is not possible to solve $\frac{d\bar{u}}{dB_o} = 0$ analytically but a rather good estimation can be obtained in linking asymptotic expressions, method which is not indispensable here but will be needed for the case of the Couette flow.

$$\frac{1}{3}\left(H^2\frac{E_o}{B_o} - \frac{a^2}{\eta}\frac{\partial p}{\partial x}\right) = \frac{E_o}{B_o} - \frac{1}{\sigma B_o^2}\frac{\partial p}{\partial x} \quad (24)$$

The three solutions of the third degree equation (24) are $B_o = \frac{1}{\sigma E_o}\frac{\partial p}{\partial x}$, which leads to $\bar{u} = 0$, and $B_o a\sqrt{\sigma/\eta} = \pm\sqrt{3}$.

Then, the optimal Hartmann number is $\mathrm{H}_{\mathrm{opt}_0} = \sqrt{3}$, value in good agreement (roughly 8 %) with the exact value obtained in solving numerically equation $\frac{d\bar{u}}{dB_o} = 0$ with leads to $\mathrm{H}_{\mathrm{opt}_0} \approx 1{,}606$. Thus, we deduce $K_{\mathrm{opt}_0} = \frac{1}{1 - \mathrm{th}(\mathrm{H}_{\mathrm{opt}_0})/\mathrm{H}_{\mathrm{opt}_0}} \approx 2{,}35$ hence $J_{\mathrm{opt}} = 0{,}57\,\sigma E_o$ and, if there is no pressure gradient:

$$\bar{u}_{\mathrm{opt}} = \bar{u}_{\max} = \frac{1}{K_{\mathrm{opt}_0}}\frac{E_o}{B_{opt}} \approx 0.43\frac{E_o}{B_{opt}} = 0.13\,\Omega\,\frac{a}{b}\sqrt{\frac{\sigma}{\eta}}$$

Contrary to $B_o$, $\bar{u}$ always increases with $E_0$ which can be seen as a limit for the induced field $\mathbf{v} \times \mathbf{B}$.

The previous optimal Hartmann number obtained is very interesting for a propulsion nozzle of a ship. Following [4], we consider a propulsion nozzle in a control volume V delimited by the upstream area $S_\infty$, $S_{ext}$ which is the lateral area, and a downstream area $S_{down}$ (refer to Fig. 5 below). The area $S_0$ is the upstream limit of the water which goes through the nozzle, and $p_\infty$ is the ambient pressure of the water. The volume $V$ is taken with $S_\infty$ and $S_{ext}$ enough far from the nozzle to be pierced by a uniform flow at $u_\infty$. The integral momentum equation for the fluid, allows obtaining the thrust $F$ for a propulsion nozzle in a steady motion with the Newton's third law (refer to [4] for further details):

$$-F = |F| = \iint_{S_{down}} \rho u^2 dS - \iint_{S_{ext}} \rho u^2 dS \quad (25.a)$$

Convert [4] notices a wrong one-dimensional simplification of the integral momentum equation in [8], sometimes quoted.

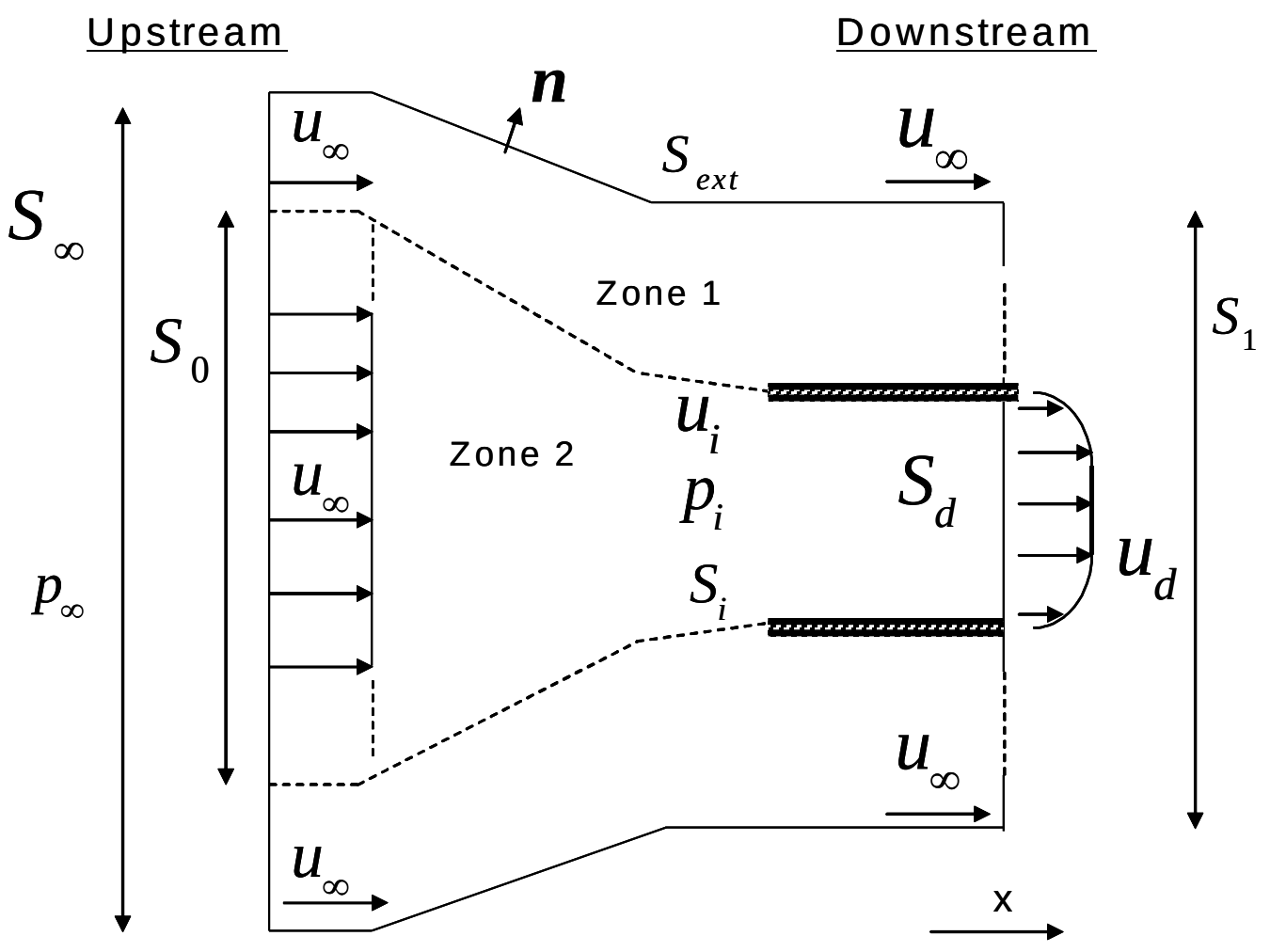

**Fig. 5. Propulsion nozzle**

Here, the velocity profiles are not supposed uniform, which lead to use kinetic energy coefficients $\beta = \frac{1}{S}\int_S \left(\frac{u}{\bar{u}}\right)^2 dz$ equal by definition to unity for a uniform velocity profil:

$$-F = |F| = \dot{m}(\beta_{down}\,\bar{u}_{down} - \beta_\infty u_\infty) \qquad (25\text{ b})$$

where $\dot{m} = \rho S_\infty \bar{u}_\infty = \rho S_{down}\,\bar{u}_{down}$ is the mass flow.

For a Hartmann velocity profiles, $\beta$ is found to be:

$$\beta = \frac{\mathrm{H}}{4}\,\frac{\mathrm{sh}(2\mathrm{H}) - 8\,\mathrm{ch}(\mathrm{H})\,\mathrm{sh}(\mathrm{H}) + 2\mathrm{H}\left[1 + 2\,\mathrm{ch}^2(\mathrm{H})\right]}{\mathrm{H}^2\,\mathrm{ch}^2(\mathrm{H}) - \mathrm{H}\,\mathrm{sh}(2\mathrm{H}) + \mathrm{sh}^2(\mathrm{H})}$$

which decreases with increasing $\mathrm{H}$ and allows recovering the classical value for $\mathrm{H} \ll 1$ : $\beta \to 6/5 = 1.2$ .

Moreover, to use the usual Bernoulli's theorem between $S_\infty$ and $S_{down}$ in taking a velocity profile into account, coefficients $\alpha$ are needed, which allows writing:

$$\alpha_\infty \frac{1}{2}\rho\bar{u}_\infty^2 + \Delta p_{MHD} = \alpha_{down}\frac{1}{2}\rho\bar{u}_{down}^2 + J_{vis} \qquad (26)$$

where $\Delta p_{MHD} = \int_L \mathbf{J}\times\mathbf{B}\cdot\mathbf{dx} = \sigma E_o B_o L - \sigma\bar{u}_d B_o^2 L$ , $J_{vis}$ is the viscous lost and $\alpha = \frac{1}{S}\int_S \left(\frac{u}{\bar{u}}\right)^3 dz$ is the momentum flow coefficient.

For a Hartmann flow, $\alpha$ is equal to $\frac{\mathrm{H}^2}{12}\frac{N}{D}$ with:

$$N = 3\mathrm{H}\,\mathrm{ch}(3\mathrm{H}) - \mathrm{sh}(3\mathrm{H}) + \mathrm{ch}(\mathrm{H})\cdot\left[27\,\mathrm{H} + 9\mathrm{sh}(2\mathrm{H})\right] - \mathrm{sh}(\mathrm{H})\cdot\left[27 + 18\,\mathrm{ch}(2\mathrm{H})\right]$$

and:

$$D = \mathrm{H}\,\mathrm{ch}(\mathrm{H})\cdot\left[\mathrm{H}^2\mathrm{ch}^2(\mathrm{H}) + 3\mathrm{sh}^2(\mathrm{H})\right] - \mathrm{sh}(\mathrm{H})\cdot\left[\mathrm{sh}^2(\mathrm{H}) + 3\mathrm{H}^2\mathrm{ch}^2(\mathrm{H})\right]$$

Thus $\frac{\mathrm{H}^2}{12}\frac{N}{D}\xrightarrow[\mathrm{H}\to 0]{} 54/35 \approx 1.5$ . It is interesting to remind that at large $\mathrm{H}$ , the velocity profile is almost uniform, which is linked with the limits of $\alpha$ and $\beta$ : $\alpha,\beta \xrightarrow[\mathrm{H}\to\infty]{} 1$ . In steady motion, the thrust is made up for the extern drag $T$ of the ship linked with the nozzle, and then, with a drag coefficient $C_x$ and a wetted area $S_w$ :

$$-F = T$$

$$\Leftrightarrow\ \rho S_{down}\,\bar{u}_{down}\,(\beta_{down}\,\bar{u}_{down} - \beta_\infty\bar{u}_\infty) = \frac{1}{2}S_w C_x \bar{u}_\infty^2 \qquad (27)$$

from which $\bar{u}_{down}$ can be obtained and allows writing Eq. (26) in function of $\bar{u}_\infty$ . An expression of viscous lost is needed to solve Eq. (26). In turbulent flows, an empiric expression can be used, but for a Hartmann flow, a theoretical expression can be found from [10] $\frac{\partial p}{\partial x} = \frac{\eta}{a^2}\mathrm{H}^2\frac{E_o}{B_o} - \frac{\eta}{a^2}\frac{\mathrm{H}^3}{\mathrm{H} - \mathrm{th}(\mathrm{H})}\bar{u}$ .

The pressure increases with the first term, equal to $\sigma E_o B_o$ , and is limited by the second one, which is the sum of the viscous lost $\eta\Delta\cdot\mathbf{v}$ in Eq. (7), and the induced electromagnetic force $\mathbf{j}_{ind}\times\mathbf{B} = (-\sigma\mathbf{v}\times\mathbf{B})\times\mathbf{B}$ . Then, the purely viscous lost $J_{vis}$ is $J_{vis} = \frac{\eta}{a^2}\frac{\mathrm{H}^3}{\mathrm{H} - \mathrm{th}(\mathrm{H})}\bar{u}_d - \sigma B_o^2\bar{u}_d = \frac{\eta}{a^2}\frac{\mathrm{H}^2\,\mathrm{th}(\mathrm{H})}{\mathrm{H} - \mathrm{th}(\mathrm{H})}\bar{u}_d$ .

With this latter expression, Eq. (26) can be solved, and $\bar{u}_\infty$ is obtained. With the optimal quantities obtained upper, a first estimation of the maximum for $\bar{u}_\infty$ has been calculated, but this estimation is correct only asymptotically, when the flow is not too far from the purely Hartmann flow. Then, to improve this estimation, as $\frac{d\bar{u}_\infty}{dB_o} = 0$ cannot be solved analytically, we have established a condition on the Hartmann number $\mathrm{H}_{opt}$ and the load factor $K_{opt}$ to reach the maximum of $\bar{u}_\infty$ :

$$K_{opt}\,\mathrm{H}_{opt} = 2\left(\mathrm{H}_{opt} - \mathrm{H}_{opt_0}\right) + K_{opt_0}\mathrm{H}_{opt_0} \qquad (28)$$

Then, if (28) is verified, the mean velocity is maximum and is found to be equal to :

$$\frac{\bar{u}_\infty}{\tilde{u}} = \varpi\,\frac{4\mathrm{H}_{opt}\left(\mathrm{H}_{opt} - \mathrm{H}_{opt_0}\right) + K_{opt_0}H_{opt_0} - 2\mathrm{H}_{opt}^3\left[\mathrm{H}_{opt} - \mathrm{th}\left(\mathrm{H}_{opt}\right)\right]^{-1}}{\alpha_d\varpi^2 - \alpha_\infty}$$

where $\tilde{u} = \frac{\eta L}{\rho a^2}\frac{S_d}{S_{MHD}}$ is a characteristic velocity, and $\varpi = \frac{1}{2}\left(\frac{\beta_\infty}{\beta_d} + \sqrt{\left(\frac{\beta_\infty}{\beta_d}\right)^2 + \frac{2S_w C_x}{\beta_d S_d}}\right)$. This analytic expression of the maximum mean velocity $\bar{u}_\infty$ , based on (28), is plotted on Fig. 6, and compared to exact maxima (points) of velocity.

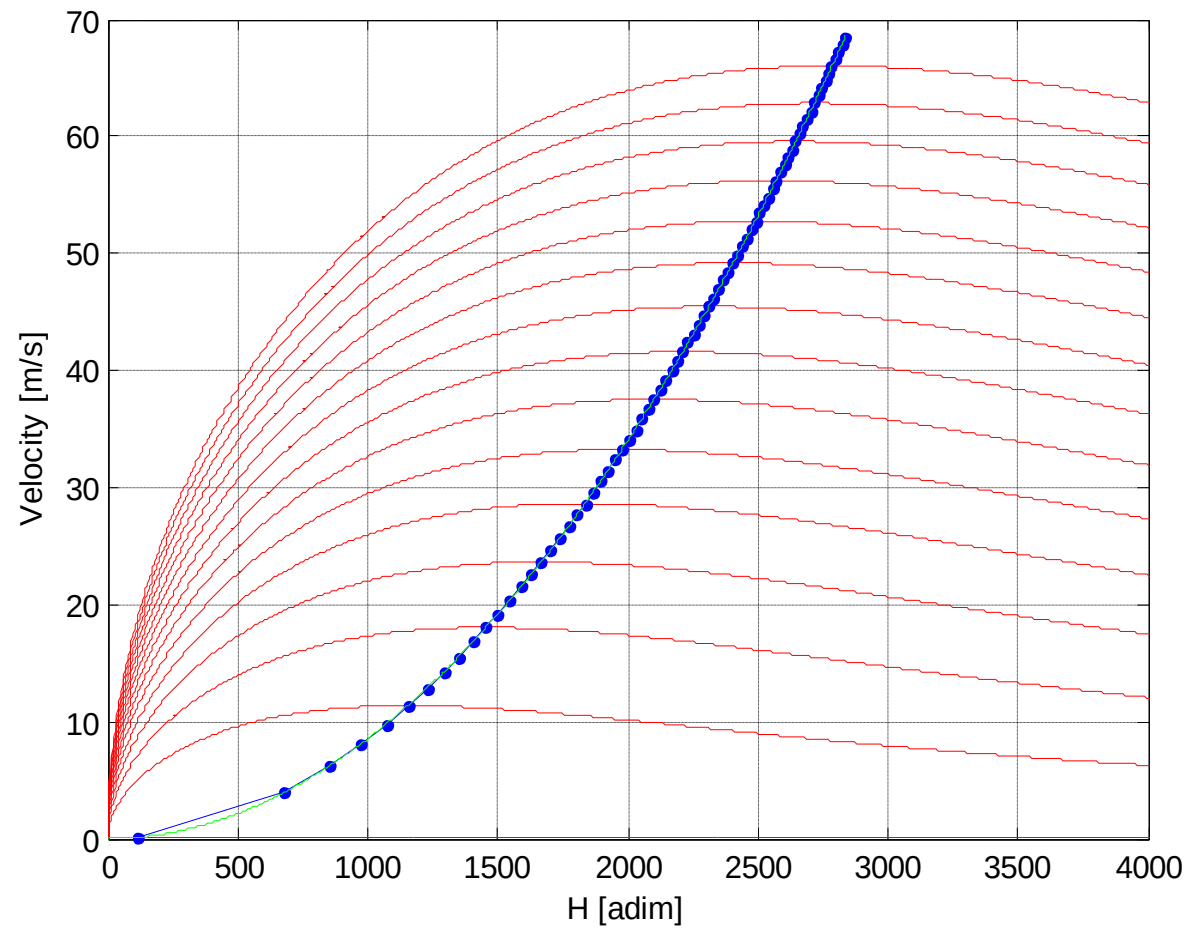

**Fig. 6. Comparison of the exact velocity maximum $\bar{u}_\infty$ (points) and the fitting (velocity plotted for different electric field)**

To conclude this section on propulsion nozzles, one can notice that in the previous study, the electric field was assumed enough weak to avoid electrolysis, but in general, this is not the case, and a coupling between electrolysis and hydrodynamics have to be taken into account (see e.g. [5] for details). However, under hypothesis, it is possible to solve a flow with electrolysis of the fluid as we will see with the study of the Couette flow in the following part.

## 2. COUETTE PROBLEM

In this section, previous solution will be extended to a problem generally referred as Couette flow (see Fig. 6): with the previous problem, the lower wall stays stationary but the other, at $z = L$ is moving with a constant velocity $u(L) = u_w$ .

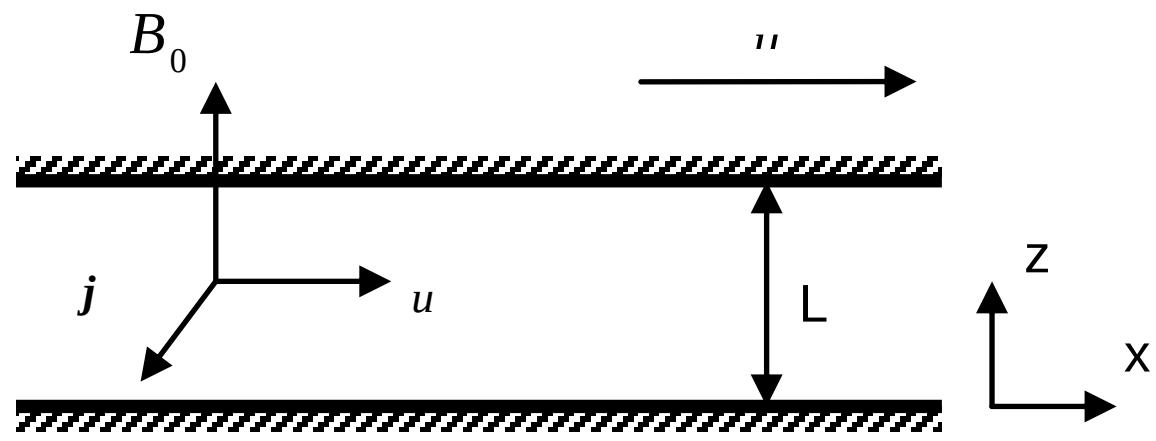

**Fig. 6. Magnetohydrodynamic Couette flow**

Since this problem is quite similar to the previous one, with the same hypothesis, one can write $\mathbf{v} = (u_x, 0, 0)$, $\mathbf{j} = (0, j_y, 0)$, $\mathbf{B} = (b_x, 0, B_o)$, $\mathbf{E} = (0, E_o, 0)$ where $E_o$ and $B_o$ are constants.

### 2.1. Classical analytical solution

As in the previous section, the heat-transfer and fluid motion equations are uncoupled and it is possible to solve equations separately. Moreover the equations are linear and, following e.g. [3], the solution for u is found to be (see fig. 7) with the boundary conditions $u(0) = 0$, and $u(L) = u_w$,

$$U = \frac{u_x}{u_w} = K + \frac{(1-K)\,\mathrm{sh}(\mathrm{ZH}) - K\,\mathrm{sh}[\mathrm{H}(1-Z)]}{\mathrm{sh}(\mathrm{H})} \tag{29}$$

$$\frac{\bar{u}}{u_w} = K + (1-2K)\frac{1}{\mathrm{H}}\,\mathrm{th}\left(\frac{\mathrm{H}}{2}\right) \tag{30}$$

where $Z = \dfrac{z}{L}$, $\mathrm{H} = aB_o\sqrt{\dfrac{\sigma}{\eta}}$ and $K = \dfrac{E_o}{u_w B_o}$.

Then, as in the previous section, then following relationship are retrieved:

$$\tilde{j}_y = \frac{j_y}{\sigma \bar{u} B_o} = K - \frac{u_x}{\bar{u}} = K - U \tag{31}$$

$$E_o = \frac{\Omega}{2b} \tag{32}$$

$$\tilde{b}_x = \frac{b_x}{B_o \mathrm{R_m}} = \frac{\mathrm{ch}(\mathrm{H}) - \mathrm{ch}(Z\mathrm{H})}{\mathrm{H}\,\mathrm{sh}(\mathrm{H})} \tag{33}$$

with the boundary condition $b_x(L) = 0$ .

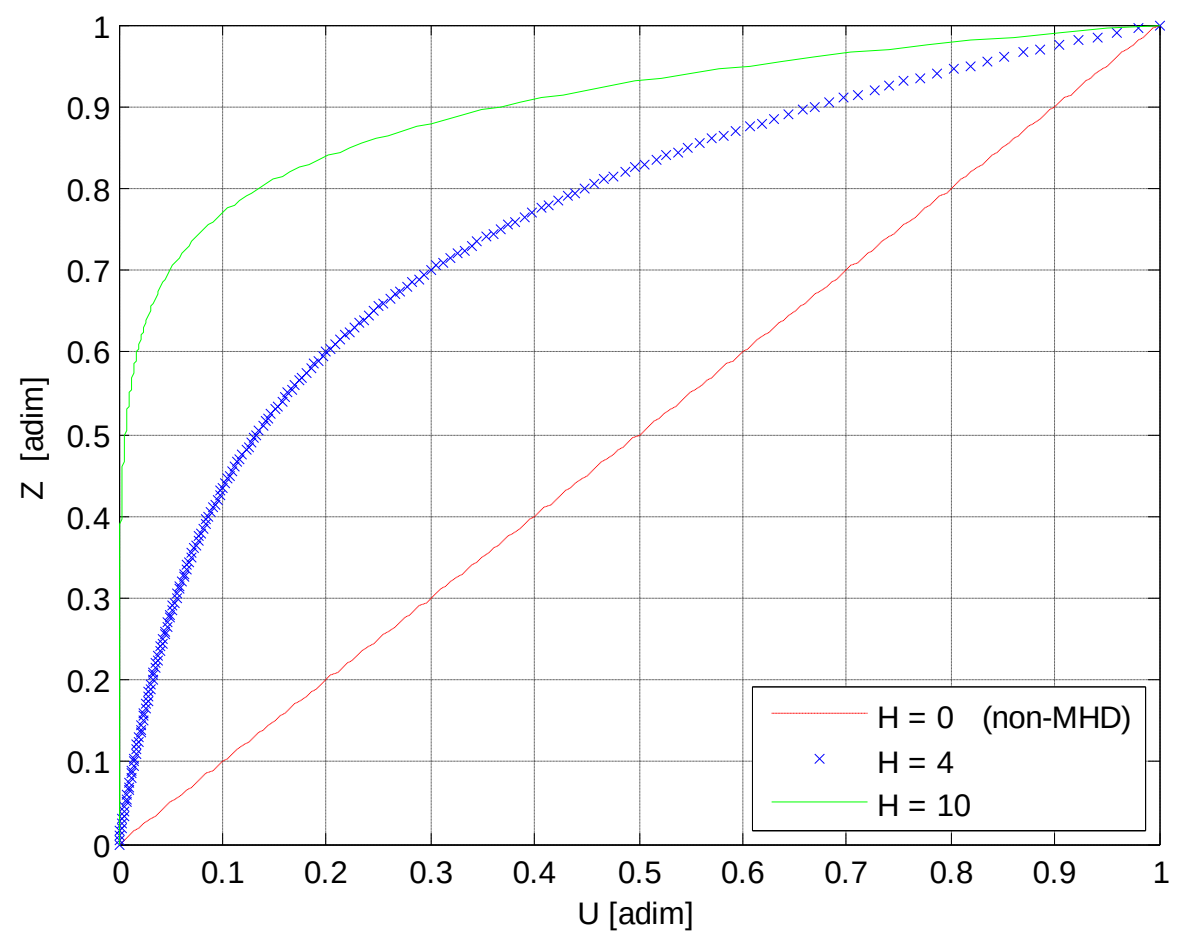

**Fig. 7. Couette flow velocity profiles ($K = 0$)**

With $e = C_v T$, the boundary condition for T is assuming to be : $T(0) = T_w$ and $T(L) = T_w$. Then, using the reduced temperature $\theta = \dfrac{C_v}{u_w^2}(T - T_w)$, the solution of (8) is (see Fig. 8): in noting $\mathrm{P_r} = \dfrac{\eta C_v}{\lambda}$ and $\tilde{r} = r\dfrac{C_v a^2}{\lambda u_w^2}$,

$$\theta_2 = \frac{P_r}{4\mathrm{sh}^3(H)}\left[\frac{K^2-4K+2}{2}\mathrm{sh}(H)+\frac{1-2K}{2}\mathrm{sh}(H(2Z-1)) + (K^2-K)[\mathrm{sh}(2ZH)-\mathrm{sh}(2H)] - \frac{(K-1)^2}{2}\mathrm{sh}(H(2Z+1)) + K(1-K)\mathrm{sh}(2H(Z-1)) + \frac{K^2}{2}[\mathrm{sh}(H(2Z-3))+\mathrm{sh}(3H)]\right] + \left[P_r\frac{1-2K}{2}+\frac{\tilde{r}(1-Z)}{2}\right]Z \tag{34}$$

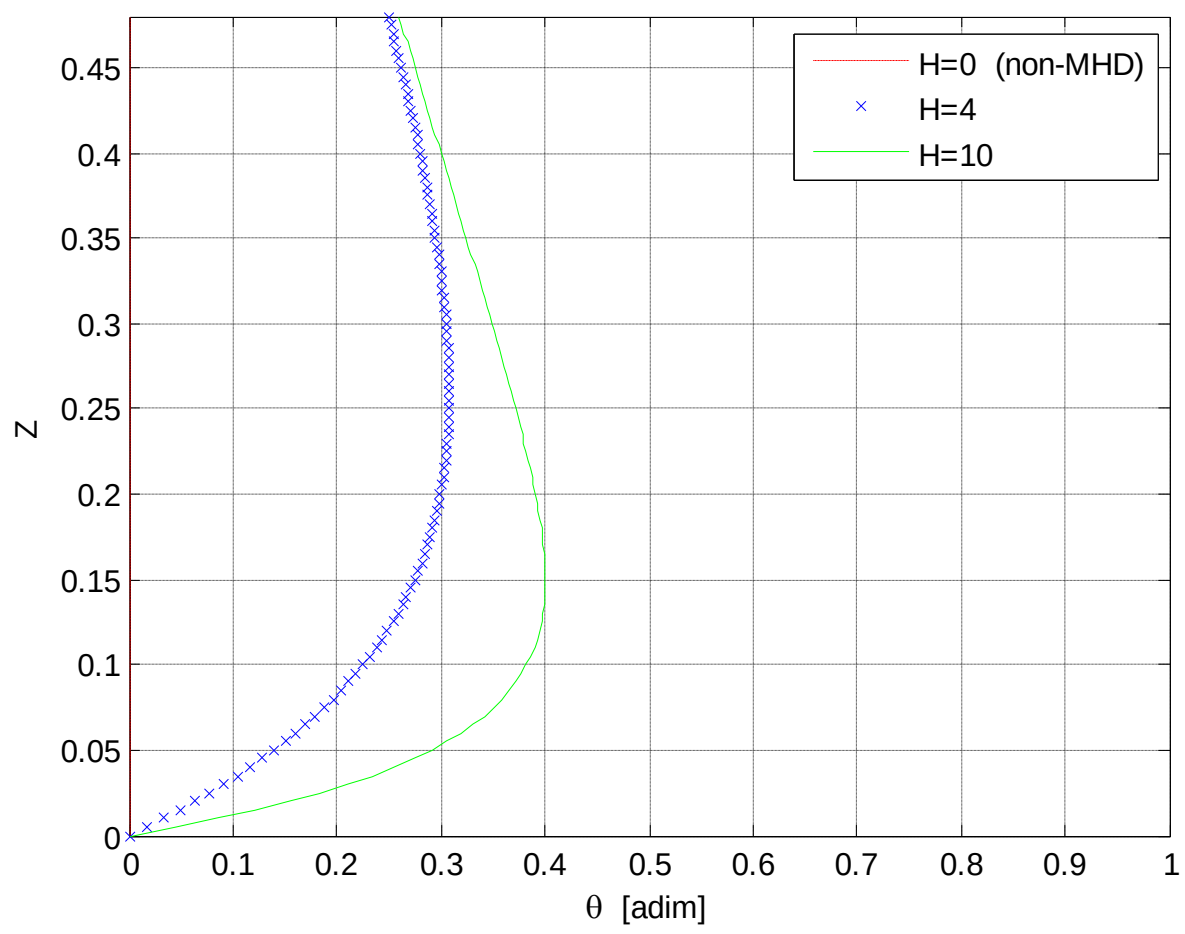

**Fig. 8. Couette flow temperature distribution ($K \cdot \bar{u} = 100$)**

Figure 8 shows clearly the large contribution of ohmic dissipation, absent in classical hydrodynamic solution, which does not allow to see the parabolic distribution of this case. A maximum can also be seen on this figure, but the equation $\frac{d\theta}{dZ}=0$ seems really difficult to solve. An estimation of the solution can be obtained with Taylor's developments near the two walls. Then, neglecting the source term $\tilde{r}$, in linking the two developments at order 1 in Z, we obtain:

$$Z_{max} = \frac{\mathrm{ch}(H)+\frac{K-1}{H}\mathrm{sh}(H)+K^2[\mathrm{ch}(H)-1]-K[2\mathrm{ch}(H)-1]}{\mathrm{ch}(H)+2K^2[\mathrm{ch}(H)-1]-2K[\mathrm{ch}(H)-1]} \tag{35}$$

The comparison of this analytic estimation with numeric solutions shows that this estimation is correct at less than 5% for the range $K > 7$ and $H < 2$.

It is also interesting to know that an analytical solution for the temperature distribution of a Couette flow has also been found when $\lambda$ is a linear function of T, but the solution has too many terms to be reproduced here.

## 2.2. Study of the Couette flow analytical solutions

First of all, as we have done for the Hartmann flow, it is interesting to study the asymptotic expressions of previous solutions:

- $H \ll 1$

$$U = \frac{u_x}{u_w} = Z = \frac{z}{L} \tag{36}$$

$$\bar{U} = \frac{\bar{u}}{u_w} = \frac{1}{2}+\left(\frac{K}{12}-\frac{1}{24}\right)H^2 + O(H^4) \tag{37}$$

$$\theta = \frac{1}{2}(P_r+\tilde{r})Z(1-Z) \tag{38}$$

- $H \gg 1$

$$U = K + 2e^{-H}[(1-K)\mathrm{sh}(ZH) - K\,\mathrm{sh}[H(1-Z)]] \tag{39}$$

$$\bar{U} = K + \frac{1-2K}{H} \tag{40}$$

As in the previous section, when $H$ is assumed to be nul, the classical solution for no applied magnetic field is recovered $\frac{\bar{u}}{u_w} = K + \frac{1-2K}{H} \xrightarrow[B_0 \to \infty]{} 0$ at large Hartmann number.

Moreover, for little value of $B_o$, $\bar{u}$ increases with $B_o$ since: $\frac{\bar{u}}{u_w} = \frac{1}{2}+\frac{K}{12}H^2 + O(B_0^2)$.

Then, a maximum exists, given by $\frac{d\bar{u}}{dB_o}=0$. This latter equation seems very difficult to solve analytically but a rather good estimation can be obtained in linking asymptotic expressions, method necessary here because $\frac{d\bar{u}}{dB_o}$ is not a function of H alone. Thus, we write:

$$\frac{1}{2}+\frac{E_o \cdot L}{12 \cdot u_w}\frac{\sigma}{\eta}B_o = \left(\frac{E_o}{u_w}+\frac{1}{L}\sqrt{\frac{\eta}{\sigma}}\right)\cdot\frac{1}{B_o} \quad \Leftrightarrow \quad KH = \frac{12-6H_{opt}}{H_{opt}^2-12} \tag{41}$$

The solution of Eq. (41) is a function of the variable $KH$, which is independent of $B_o$:

$$H_{opt} = \frac{-3+\sqrt{9+12KH(1+KH)}}{KH} = f(KH) \tag{42}$$

The comparison of this optimal Hartmann number with numeric solutions shows that this estimation is correct at less than 15% for KH > 20. One can notice also that $H_{opt} - 1/4$ is a better estimation since the gap is less than 7% for KH > 20 and decreases around 0 for large values of KH, which will be explained in linking Couette flow with Hartmann flow.

The Hartmann flow can be seen like a Couette flow with a wall velocity null. Then, the analytical solutions of the Couette flow are more general and it should be possible to recover solutions of the Hartmann flow with $u_w = 0$. We can write $L = 2a$ so $H = 2H_H$ where $H$ is the Hartmann number for the Couette flow and $H_H$ for the Hartmann flow. Hence $Z_1 = \frac{z_1}{L} \Rightarrow Z_2 = \frac{z_2}{a} = 2Z_1 - 1$ with the same notations. Then, with $u_w = 0$, Eq. (30) allows to recover Eq. (10):

$$\bar{u} = \frac{E_o}{B_o} - 2\frac{E_o}{B_o}\frac{\mathrm{th}(\mathrm{H}/2)}{\mathrm{H}} = \frac{E_o}{B_o}\left(\frac{\mathrm{H}_\mathrm{H} - \mathrm{th}(\mathrm{H}_\mathrm{H})}{\mathrm{H}_\mathrm{H}}\right) \quad (43)$$

And so, using Eq. (43), Eq. (29) allows to recover Eq. (9), which is satisfying. The same thing can be done with the temperature with a little more calculations. Besides, with (35), we have $Z_{\max} \xrightarrow[K = \frac{E_0}{u_\infty \cdot B_0} \to \infty]{} \frac{1}{2}$. Moreover, with Eq. (42), it is satisfying to check that $\mathrm{H}_{opt} = f(K\mathrm{H}) \xrightarrow[K = \frac{E_o}{u_w \cdot B_o} \to \infty]{} \mathrm{H} = 2\mathrm{H}_\mathrm{H} = 2\sqrt{3}$. Then, the asymptotic gap of roughly $\frac{1}{4}$ for large values of KH viewed upper is nothing else than the gap between the exact solution $\mathrm{H}_{\mathrm{opt}} \approx 1{,}606$ and its estimation by $\sqrt{3}$ for the Hartmann flow, i.e. $\mathrm{H}_{est} - \mathrm{H}_{exact} = 2(\mathrm{H}_{\mathrm{H}\,est} - \mathrm{H}_{\mathrm{H}\,exact}) = 2(\sqrt{3} - 1{,}606) = 0{,}252 \approx \frac{1}{4}$. Then, the estimation (42) can be corrected by $\mathrm{H}_{opt} = \frac{-3 + \sqrt{9 + 12K\mathrm{H}(1 + K\mathrm{H})}}{K\mathrm{H}} - 0{,}252$ which supplies a very good estimation of $\mathrm{H}_{\mathrm{opt}}$.

### 2.3. Extension to a solution with electrolysis

As we have seen previously, electrolysis is difficult to avoid and have to be taken into account in general. Then, with chemical reactions at electrodes, the potential is reduced of :

$$\delta\Omega_i = E_{th_i} + \eta_i \quad (44)$$

for each electrode (cathodic and anodic potential gap)

The equilibrium reduction potential $E_{th_i}$ is given by the Nernst equation, and the overpotential $\hat{\eta}_i$ is usually given by the *Butler-Volmer equation*, which is simplified here for the sake of simplicity in the case of large density of current, into the *Tafel equation* $\hat{\eta}_i = \hat{\eta}_{i0} \cdot \ln(j)$. Then, with the total equilibrium reduction potential $E_{th} = E_{th_a} + E_{th_c}$ and the total overpotential $\hat{\eta}_0 = \hat{\eta}_{0_a} + \hat{\eta}_{0_c}$, Eq. (1) can be written as: $j_y = \sigma\left(\frac{\Omega - E_{th} - \hat{\eta}_0 \cdot \ln(j_y)}{2b} - uB_o\right)$ from which we obtain, with W the Lambert W function:

$$\hat{\eta}_0 \cdot \ln(j_y) = \Omega - E_{th} - 2buB_o - \hat{\eta}_0 \cdot \mathrm{W}\left[\frac{2b}{\sigma\hat{\eta}_0} e^{\frac{\Omega - E_{th} - 2b \cdot uB_o}{\hat{\eta}_0}}\right] \quad (45)$$

To do an analytic study, we need a more simple expression of Eq. (45) and so, we assume that the ratio $\xi$ between the induced potential $2b\,uB_o$ and the total overpotential verify $\xi = \frac{2b\,uB_o}{\hat{\eta}_0} << 1$, and then, putting $\omega = \mathrm{W}\left(\frac{2b}{\sigma\hat{\eta}_0} e^{\frac{\Omega - E_{th}}{\hat{\eta}_0}}\right)$:

$$j_y = e^{\frac{\Omega - E_{th}}{\hat{\eta}_0} - \omega}\left[1 - \frac{\xi}{1 + \omega}\right] + O(\xi^2) \quad (46)$$

Then, in writing Eq. (7), we can notice that previous equations, without electrolysis, can be recovered with the variables:

- $B_o^* = B_o\sqrt{\frac{2b}{\sigma\hat{\eta}_0(1+\omega)} e^{\frac{\Omega - E_{th}}{\hat{\eta}_0} - \omega}} \rightarrow \begin{cases} B_o & \text{if } \hat{\eta}_0 \to 0 \\ 0 & \text{if } \hat{\eta}_0 \to \infty \end{cases}$;
- $E_o^* = \frac{1}{\sigma} e^{\frac{\Omega - E_{th}}{\hat{\eta}_0} - \omega} \rightarrow \begin{cases} \frac{\Omega - E_{th}}{2b} & si\ \hat{\eta}_0 \to 0 \\ 1/\sigma & si\ \hat{\eta}_0 \to \infty \end{cases}$.

It is interesting to notice that the hypothesis $\xi << 1$ is verified if $\xi_{\max} = \frac{2b\,u_{\max}B_o}{\hat{\eta}_0} << 1$. Then, for a Hartmann flow without pressure gradient for instance, in noting $u_{\max} = \frac{E_o}{B_o} g(H)$ where $g(H)$ is equal to $\frac{H^2}{2}$ at small H and 1 at large H, we have: $\xi_{\max} = \frac{2b}{\hat{\eta}_0}\frac{E_o}{K} g(H)$ which shows that previous solutions are corrects at large K at least, whatever the value of H, as we can see on Fig.9.

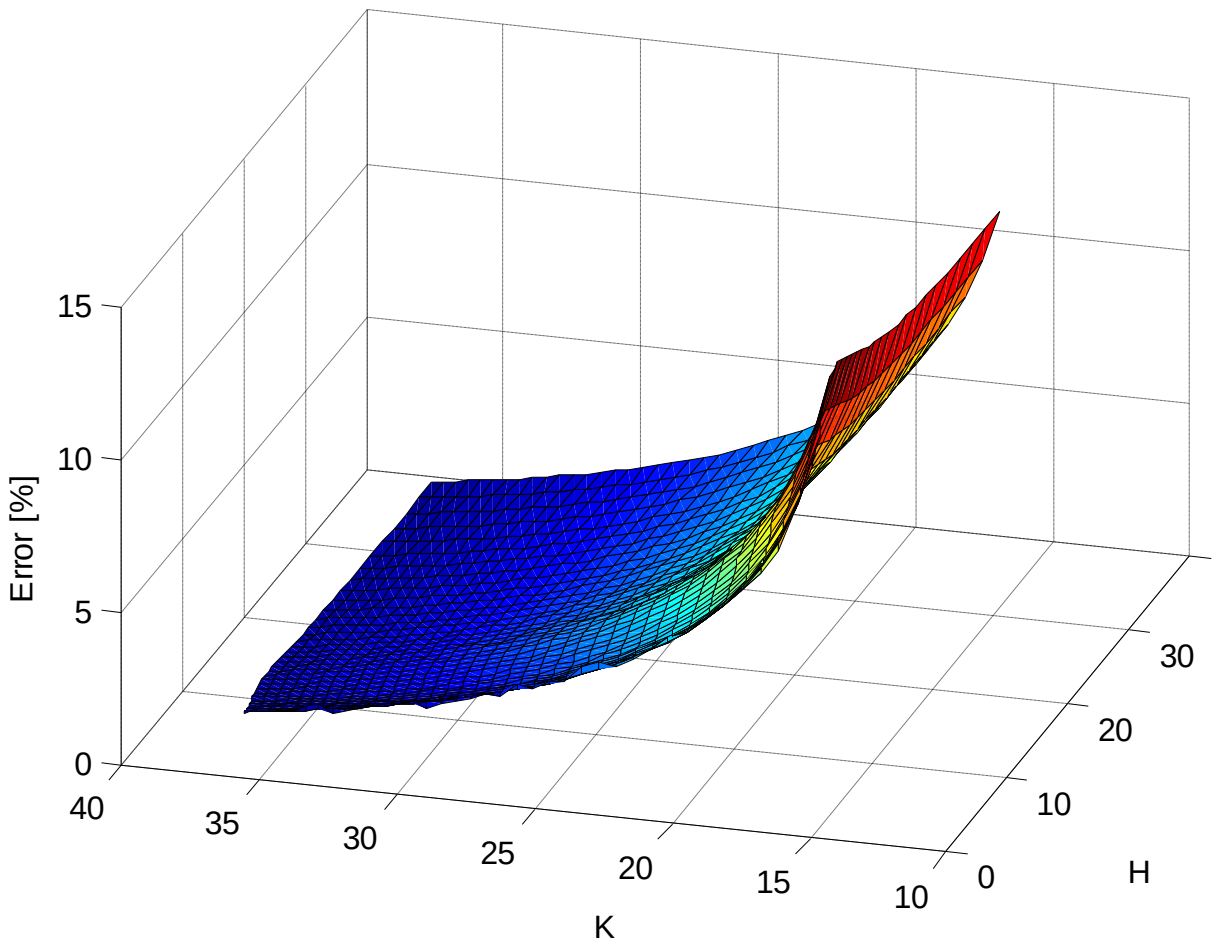

**Fig. 9. Error between the exact solution and previous analytic solutions for a Hartmann flow with electrolysis**

To conclude, with these new electromagnetic variables, if $\xi_{max} << 1$, all the results exposed above for the Couette flow, and so for the Hartmann flow also, are still correct when there is an electrolysis of the fluid.

### 2.3. Extension to another configuration

It is interesting to notice that a permutation between the electrical field and the magnetic field in the problem considered above create the same flow in the opposite direction. Then, previous solutions are still true when the velocity profile of the Couette flow, and then the Hartmann flow, is studied between two infinite plane electrodes. This property may be useful in a two-dimensional numerical simulation

### 2.4. Extension to a transient Couette flow

Some authors have studied the problem of a transient Couette flow, where the upper wall is impulsively moved. A quite simple solution can be found in the case of a fluid with a magnetic Prandtl number $P_m = \frac{\nu}{D_{mag}} = \frac{R_m}{R_e} = \mu\sigma\nu$ equal to 1 because this value of $P_m$ allows to obtain uncoupled equations. Following [3], we take the initial values and boundary conditions $\tilde{b}_x(0,Z) = 0$, $\tilde{b}_x(\tilde{t},0) = 0$, $\tilde{b}_x(\tilde{t},1) = 0$, $U(0,Z) = 0$, $U(\tilde{t},0) = 0$, $U(\tilde{t},1) = 1$ where we use these reduced variables: $U = \frac{u_x}{u_w}$, $\tilde{t} = t\frac{\eta}{\rho L^2}$, $Z = \frac{z}{L}$, $\tilde{b}_x = \frac{b_x}{B_o R_m}$ and $R_m = \frac{L\,u_w}{D_{mag}}$.

Then, the solutions, plotted on Fig. 10, are (see [3] for further details):

$$U(\tilde{t},Z) = \mathrm{ch}\left[\frac{H(Z-1)}{2}\right]\left[\frac{\mathrm{sh}(ZH)}{\mathrm{sh}(H)} + 2\pi\sum_{n=1}^{\infty}\frac{(-1)^n n\cdot\sin(n\pi Z)}{\frac{H^2}{4}+(n\pi)^2}e^{-\left[\frac{H^2}{4}+(n\pi)^2\right]\cdot\tilde{t}}\right] \quad (47)$$

$$\tilde{b}_x(\tilde{t},Z) = U(\tilde{t},Z)\cdot\mathrm{th}\left[\frac{H(Z-1)}{2}\right] \quad (48)$$

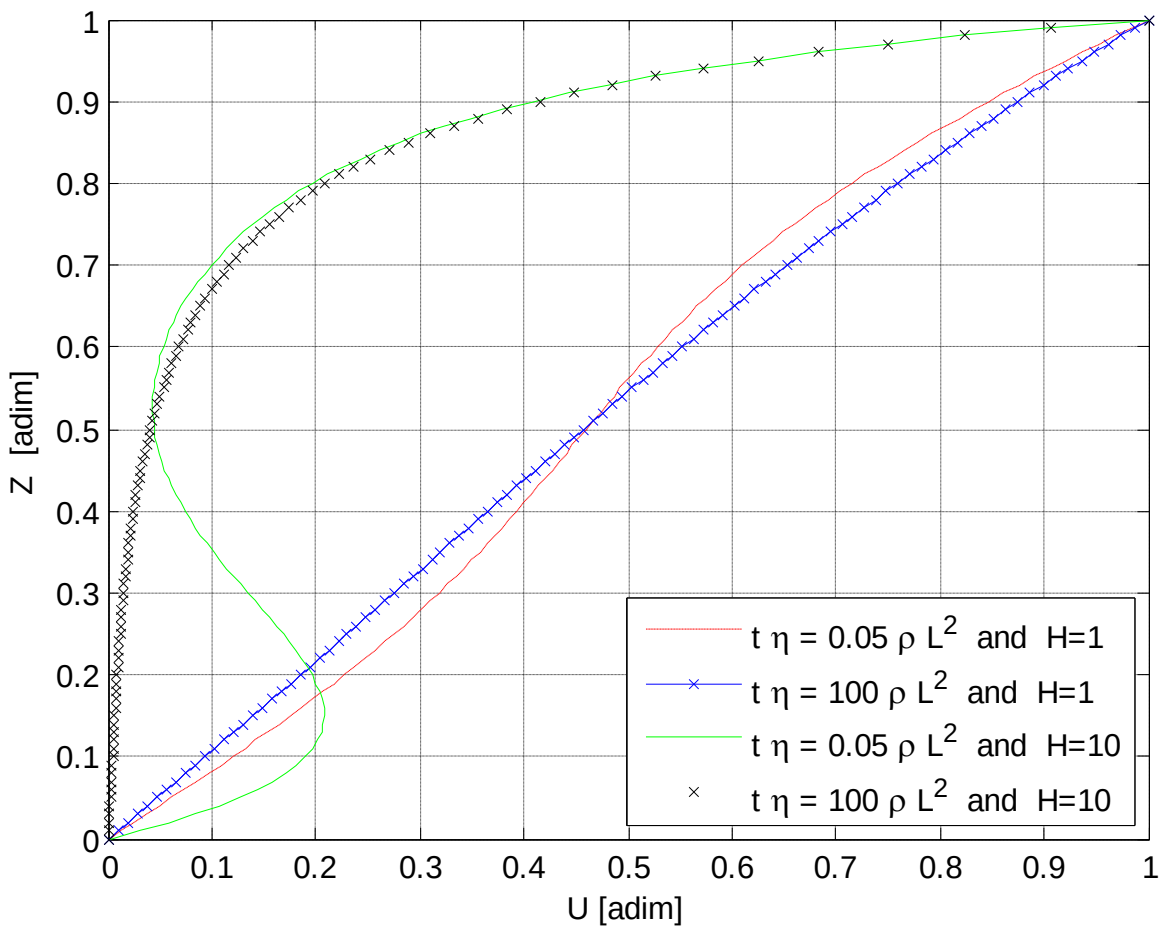

**Fig. 10. Velocity profiles for a transient Couette flow at different times for two different Hartmann numbers**

## 3. RAYLEIGH PROBLEM

The Rayleigh problem, also called first Stoke's problem, is a transient problem in which a magnetic field, assumed uniform in space and constant in time, is applied normal to the surface of an impulsively moved half plane (see Fig. 11). This problem is interesting because its solution can be obtained in closed form, so that the nature of magnetohydrodynamic boundary-layer can be inferred.

Here, because of the transient nature of the problem, it is easiest to work with the well-known *induction equation*:

$$\frac{D\boldsymbol{B}}{Dt} = (\boldsymbol{B}\cdot\nabla)\boldsymbol{v} + D_{mag}\cdot\Delta\boldsymbol{B} \quad (49)$$

which is obtained directly from the Maxwell's equation and the Ohm's law.

Then, the governing equations of the problem are (7) and (49), that is to say:

$$\rho\frac{\partial u}{\partial t} = \frac{B_o}{\mu}\frac{\partial b_x}{\partial y} + \eta\cdot\frac{\partial^2 u}{\partial y^2} \quad (50)$$

$$\frac{\partial b_x}{\partial t} = B_o\frac{\partial u}{\partial y} + D_{mag}\frac{\partial^2 b_x}{\partial y^2} \quad (51)$$

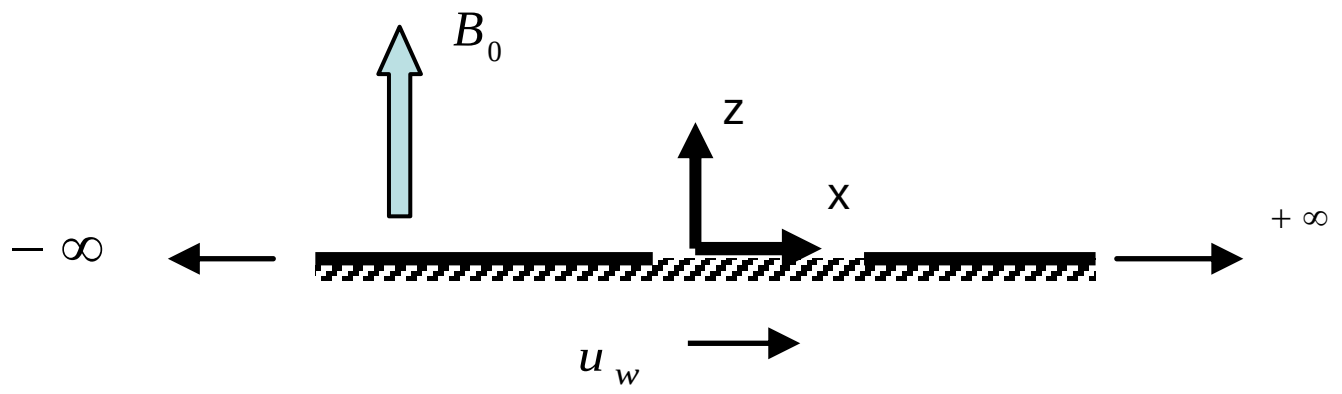

**Fig. 11. Rayleigh problem configuration**

Following [3], initial values and boundaries conditions are as follows: $u(y,0) = 0$, $b_x(y,0) = 0$, $u(0,t) = u_o$, $b_x(0,t) = 0$ with $u$ and $b_x$ being bounded when $y \to \infty$.

As in the previous section, the heat-transfer and fluid motion equations are uncoupled and it is possible to solve equations separately. However, the equations (50) and (51) show that the velocity and the induced magnetic field are solutions of a partial differential system of two equations coupled.

A general expression of the shear stress on the wall can be obtained, see [3]:

$$\tau_w(t) = \frac{\eta \cdot u_0}{\sqrt{\nu}}\left[\frac{e^{-\gamma \cdot t}}{\sqrt{\pi t}} + \sqrt{\gamma} \cdot \mathrm{erf}\left(\sqrt{\gamma \cdot t}\right)\right] \underset{t \to \infty}{\longrightarrow} \frac{\sqrt{\mathrm{P_m}}}{1+\sqrt{\mathrm{P_m}}} \qquad (52)$$

where $\gamma = \dfrac{\sigma \cdot B_o^2}{\rho\left(1+\sqrt{\mathrm{P_m}}\right)^2}$ .

It is also possible to obtain a general solution for the steady motion because in this case, (50) and (51) can be written as:

$$0 = \frac{B_o}{\mu}\frac{\partial b_x}{\partial z} + \eta \cdot \frac{\partial^2 u}{\partial z^2} \qquad (53)$$

$$0 = B_o\frac{\partial u}{\partial z} + D_{mag}\frac{\partial^2 b_x}{\partial z^2} \qquad (54)$$

which can be solved:

$$U = \frac{u}{u_0} = e^{-\frac{y}{L}} \qquad (55)$$

$$b_x = \sqrt{\sigma\eta} \cdot u_0 \mu (U - 1) \qquad (56)$$

where $\mathrm{L} = \dfrac{1}{B_o}\sqrt{\dfrac{\eta}{\sigma}}$ is a characteristic length of the problem.

However, no general solution has been found for the Rayleigh problem, and hypotheses have to be made to go further. Various approximations can be made (see references [6] and [7]) and we choose here to assume $\mathrm{R_m} << \mathrm{R_e}$ , and then, with the aid of the Laplace transform, following [3], we have (see Fig.12):

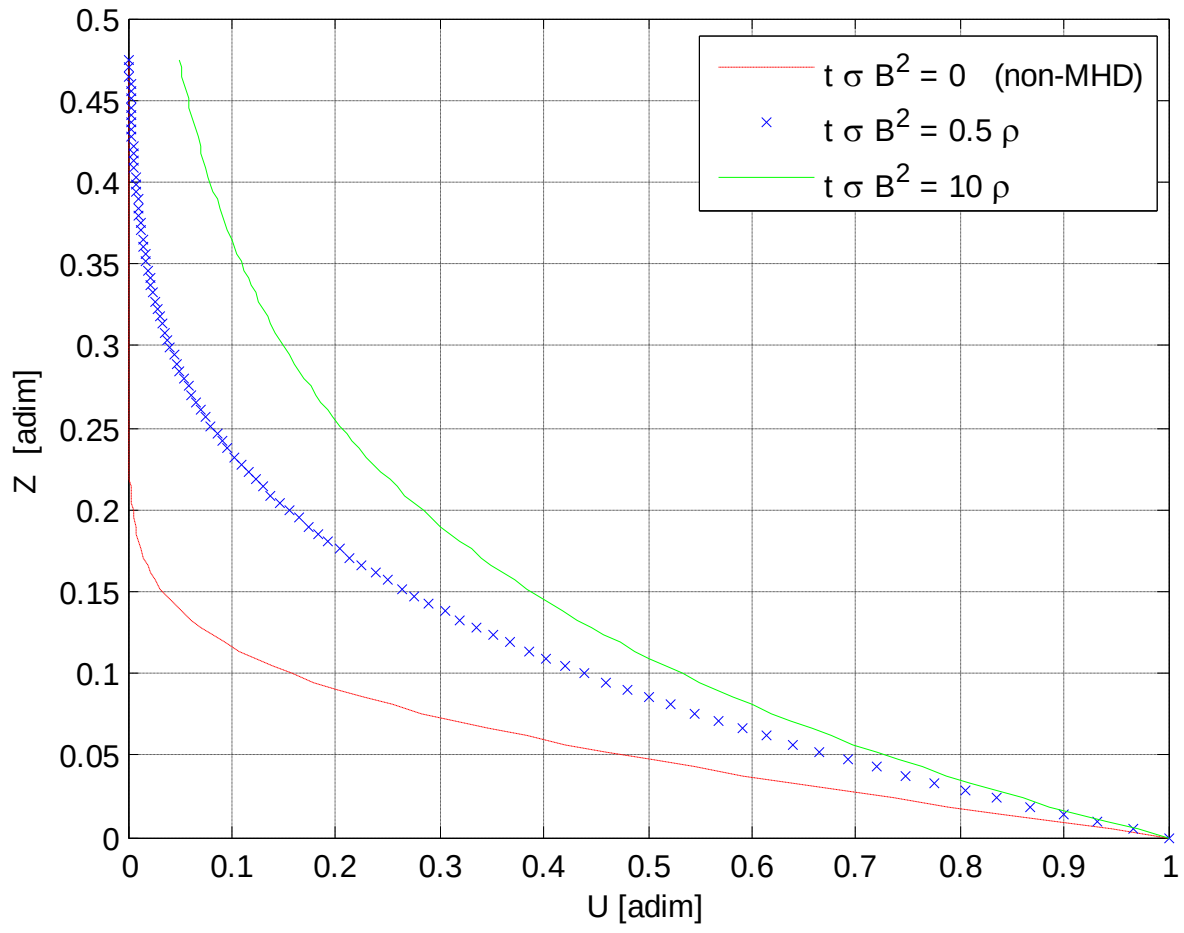

**Fig. 12. Velocity profiles in the Rayleigh problem**

$$U = \frac{u_x}{u_0} = \frac{1}{2}\left[e^{-\frac{y}{L}} \cdot \mathrm{erfc}\left(\frac{y}{2\sqrt{\nu t}} - \sqrt{\frac{\sigma \cdot B_o^2}{\rho}t}\right) + e^{\frac{y}{L}} \cdot \mathrm{erfc}\left(\frac{y}{2\sqrt{\nu t}} + \sqrt{\frac{\sigma \cdot B_o^2}{\rho}t}\right)\right] \qquad (57)$$

The magnetic field increases the time needed to reach a given value of the velocity at any point of the flow. The asymptotic expression of Eq. (57) allows to recover the classical solution of this problem:

$$H << 1: U = \frac{u}{u_0} = \mathrm{erfc}\left(\frac{\mathrm{y}}{2\sqrt{\nu t}}\right) \qquad (58)$$

The study of the temperature distribution is a boundary layer problem quite difficult to solve analytically, which has been already studied by many authors. This aspect has not been studied in this work (for details, see reference [10]).

## CONCLUSION

In the present paper, various test cases for MHD problems have been analytically investigated and physically discussed. The presented work is the starting point of a more general study which aims at validating some engineering numerical tools that can be used to model multi-physic problems.

Future publication will present a numerical procedure, based on the finite element method, which will be used to study more complex situations in MHD flows. Validation of the finite element procedure will be performed by a comparison between the numerical computations and analytical calculations on the test cases developed in the present paper. Industrial perspective of this R&D program involves an investigation of flow control with MHD-based techniques.